\documentclass[%
  reprint,superscriptaddress,amsmath,amssymb,
  aps,nofootinbib]{revtex4-1}

\usepackage{graphicx}
\usepackage[T1]{fontenc}
\usepackage{dcolumn}
\usepackage{bm}
\usepackage{xcolor}
\usepackage{epsfig}
\usepackage{float}
\usepackage{siunitx}
\usepackage{subcaption}
\usepackage{caption}

\usepackage[colorlinks=true,urlcolor=brown,linkcolor=blue,citecolor=magenta]{hyperref}

\makeatletter
\g@addto@macro\normalsize{
  \setlength\abovedisplayskip{6pt plus 2pt minus 2pt}
  \setlength\belowdisplayskip{6pt plus 2pt minus 2pt}
  \setlength\abovedisplayshortskip{4pt plus 2pt minus 2pt}
  \setlength\belowdisplayshortskip{4pt plus 2pt minus 2pt}
}
\makeatother

\newcommand{\Msun}{\,{\rm M_\odot}}
\newcommand{\Rsun}{\,R_\odot}
\newcommand{\kms}{\,{\rm km\,s^{-1}}}
\newcommand{\au}{\,{\rm au}}
\newcommand{\fdm}{f_{\rm DM}}
\newcommand{\flim}{f_{\rm lim}}
\newcommand{\Rninety}{R_{90}}
\newcommand{\lnL}{\ln\Lambda}
\newcommand{\tdis}{t_{\rm dis}}
\newcommand{\tcat}{t_{\rm cat}}
\newcommand{\tage}{t_{\rm age}}
\newcommand{\Psurv}{\mathcal{P}_{\rm surv}}
\newcommand{\Npred}{N_{\rm pred}}
\newcommand{\Nobs}{N_{\rm obs}}
\newcommand{\PR}{\mathcal{P}_{\mathcal{R}}}
\newcommand{\Mtd}{M_{\rm 2D}}
\newcommand{\dmin}{{\delta_{\chi}^{\rm min}}}

\begin{document}

\title{Wide binaries in ultra-faint dwarf galaxies as a probe of extended dark objects}

\author{María Olalla Olea-Romacho}
\email{maria.olalla.olea_romacho@kcl.ac.uk}

\affiliation{
Theoretical Particle Physics and Cosmology,
King's College London,
Strand, London WC2R 2LS, United Kingdom
}

\begin{abstract}
Wide stellar binaries are sensitive dynamical probes of dark-matter substructure. We generalize existing point-mass disruption constraints to spherical extended dark objects and apply the resulting framework to the population of wide-binary candidates identified in the ultra-faint dwarf galaxy Boötes~I. We derive $95\%$ confidence limits on the dark-matter fraction as a
function of the perturber mass $M$, radius $R_{90}$ enclosing $90\%$ of
the mass, and density profile, considering Navarro-Frenk-White, $\rho\propto r^{-3/2}$,
and $\rho\propto r^{-9/4}$ models. Sufficiently compact objects converge to the point-mass limit, whereas finite size suppresses binary disruption when the perturber becomes comparable to the relevant encounter and binary scales. The constraints remain sensitive to dark-matter fractions well below unity over a broad region of the mass–radius plane and extend to objects substantially larger than those accessible to conventional microlensing searches. As an application, we map the $\rho\propto r^{-3/2}$ results onto ultracompact minihalos and derive an illustrative constraint on the primordial curvature power spectrum. Ultra-faint-dwarf wide binaries therefore offer a purely gravitational probe of the abundance of extended dark-matter objects and of the small-scale primordial power spectrum.
\end{abstract}

\vspace{0.7cm}

\maketitle

\section{Introduction}
\label{sec:intro}

The identity of dark matter (DM) remains unknown, and a broad class of
candidates predicts that some or all of it is bound into massive
structures. Primordial black holes (PBHs) provide one of the best-studied
examples of this idea: compact relics whose abundance has been
constrained across many decades in mass by evaporation, microlensing,
dynamical effects, accretion, and gravitational-wave searches. Together,
these constraints exclude PBHs as the totality of the dark matter over
most of the non-evaporated mass range, leaving the asteroid-mass window,
$10^{-17}\Msun \lesssim M_{\rm PBH} \lesssim 10^{-10}\Msun$, as the main
region where they could still account for all of it~\cite{Carr:2026hot}.

At higher masses, a recent study based on JWST/NIRCam observations of Bo\"otes~I provided a new dynamical constraint on massive compact halo objects (MACHOs). The analysis
identified 52 candidate wide binaries with projected separations
$7000 \lesssim s/{\rm au} \lesssim 16000$ and used their survival to
constrain dark perturbers in the ultra-faint dwarf (UFD) environment
\cite{Shariat:2025dxs}. For point-like perturbers, including PBHs, the
resulting bound limits the dark-matter fraction to below ${\sim}1\%$ for
$M \gtrsim 5\Msun$, tightening to the ${\sim}0.1\%$ level at higher
masses. The strength of this constraint is attributed to the old stellar
population of Bo\"otes~I, its extreme dark-matter domination,
with a stellar-to-dark-matter mass ratio $M_\star/M_{\rm DM} \simeq 10^{-5}$--$10^{-4}$, and the scarcity of
ordinary baryonic perturbers. However, these bounds remain sensitive to assumptions about the initial binary population, the modeling of stellar flybys, and residual contamination from chance alignments.

These results provide constraints on compact objects treated as point-like perturbers, but many well-motivated candidates
have finite extent: ultracompact minihalos (UCMHs) formed from the
collapse of enhanced primordial overdensities
\cite{Delos:2018ueo, Bringmann:2011ut, Ricotti:2009bs},
prompt cusps \cite{DelosWhite2023, Delos:2019mxl, Olea-Romacho:2025qag, Olea-Romacho:2025mxj}, boson stars~\cite{Kaup:1968zz, Visinelli:2021uve}, axion
miniclusters~\cite{Fairbairn:2017sil, Hogan:1988mp, Kolb:1993zz}, and other extended dark objects (EDOs). For these objects,
the sensitivity of a given probe depends not only on the total mass but
also on the physical size and the density profile.

The mass–size dependence is cleanly illustrated by microlensing. For a
compact lens, the relevant scale is the Einstein radius, and the signal
is governed essentially by the lens mass alone. Once the lens is
extended over a scale comparable to or larger than its Einstein radius,
finite-size effects suppress the magnification, and the constraints
weaken rapidly relative to the point-mass case
\cite{Croon:2020wpr, Croon:2020ouk}.
Caustic-crossing events in giant arcs~\cite{Kelly:2014mwa, Kelly:2017fps} extend this logic to larger
physical radii by exploiting cosmological distances and the
macro-magnification of the cluster lens: in favourable configurations
they can probe objects with sizes up to $R_{90} \lesssim 10^7\Rsun$,
where $R_{90}$ denotes the radius enclosing $90\%$ of the object's mass~\cite{Croon:2025yfj}.

Dynamical probes offer a complementary window on both compact and extended dark objects. Rather than relying on lensing magnification, they exploit the gravitational response of weakly bound or kinematically cold systems to repeated gravitational encounters~\cite{2008gady.book.....B, Penarrubia2019art, HamiltonShaunak24}. Examples include the heating and expansion of stellar populations in ultra-faint dwarfs~\cite{Brandt:2016aco, Graham:2024hah}, perturbations of stellar streams and satellite counts~\cite{Ando:2022tpj}, the disruption of ultra-wide exoplanetary systems~\cite{Pare:2026rtm}, and the survival of wide stellar binaries~\cite{1985ApJ...290...15B, 1987ApJ...312..367W}, both in the Milky Way~\cite{Ramirez:2022mys} and in dwarf galaxies~\cite{Shariat:2025dxs}.
In the Milky Way, wide binaries have served as dynamical probes of compact dark matter: constraints from halo-binary samples \cite{Yoo:2003fr, Quinn:2009zg, Monroy-Rodriguez:2014ula} limit MACHOs heavier than a few tens of solar masses to a subdominant fraction of the halo, although the precise limits depend on assumptions about the initial binary population~\cite{Tyler:2022rxi}, and Gaia catalogues have recently extended such analyses to extended substructure~\cite{Ramirez:2022mys}. In the Galactic environment, however, the local dark-matter density is more than an order of magnitude below that of ultra-faint dwarfs~\cite{Read:2014qva, Shariat:2025dxs}, and disruption by stellar encounters, molecular clouds, and the Galactic tide~\cite{Jiang:2009ax} dominates the total perturbation rate, so any dark-matter signal must be extracted from beneath a modeled baryonic foreground.  

All of these observables probe the same underlying physics: a population of massive dark objects transfers orbital energy to stars, planets, or stellar systems through gravitational scattering. UFDs provide a particularly promising environment for these dynamical tests, and upcoming surveys will greatly expand the available sample of faint systems. The Vera C.~Rubin Observatory will substantially enlarge the census of Milky Way satellites, while the Nancy Grace Roman Space Telescope will extend resolved studies of ultra-faint dwarfs beyond the Local Group~\cite{LSST:2008ijt, LSSTDESC:2025hol, LSSTDarkMatterGroup:2019mwo, Nadler:2024ims}.

In this paper, we extend the Bo\"otes~I wide-binary disruption
constraint from point masses to spherical extended perturbers. We
generalize the impulsive-heating calculation to objects with arbitrary
density profiles (Sec.~\ref{sec:formalism}), verify that the point-mass
result is recovered in the compact limit, and derive $95\%$ exclusion
limits on the dark-matter fraction in the model-agnostic $(M,\Rninety)$
plane (Sec.~\ref{sec:constraints}). We consider three benchmark profiles: a Navarro-Frenk-White (NFW)~\cite{Navarro:1995iw}, a prompt-cusp-like profile
$\rho\propto r^{-3/2}$, and a UCMH-like profile
$\rho\propto r^{-9/4}$, spanning radii
$\Rninety=10^4$--$10^7\Rsun$. As an
application, we interpret the $\rho\propto r^{-3/2}$ results in terms of
ultracompact minihalos and translate the corresponding abundance limit
into a constraint on the primordial curvature power spectrum $\PR(k)$
(Sec.~\ref{sec:ucmh_PR}). Wide binaries in ultra-faint dwarfs therefore
probe not only the abundance of dark perturbers, but also the primordial conditions that produced them.

While this work was in preparation, Ref.~\cite{Caputo:2026fuj} appeared, presenting a general Fokker--Planck framework for the evolution of binaries under stochastic perturbations, including a treatment of extended perturbers in the impulsive regime relevant for this study. The analysis presented here was developed independently and adopts a simpler description of binary evolution; the resulting constraints could therefore be revisited within the more general framework of Ref.~\cite{Caputo:2026fuj}. Our focus is instead on constraining the abundance of extended dark objects and applying these limits to ultracompact minihalos and the primordial curvature power spectrum.

\section{Wide-binary constraints from UFDs}
\label{sec:pointmass}

\subsection{The Bo\"otes~I observable}
\label{sec:observable}

Shariat et al.~\cite{Shariat:2025dxs} identify 52 wide-binary candidates in Bo\"otes~I with
projected separations $7000 \lesssim s/\au \lesssim 16000$. Their
MACHO constraint uses the subsample with $s \geq 9000\au$, where the
survey is approximately complete for binary mass ratios $q \gtrsim 0.6$, and
conservatively assumes that only half the candidates are genuine
binaries. Counting the published catalogue, 43 pairs satisfy
$9000 \leq s/\au \leq 16000$, giving $\Nobs = \lfloor 43/2 \rfloor =
21$. 

The initial population is referenced to a wide-binary fraction
$f_{\rm wb,0} = 2.5\%$ of $N_{\rm tot} \approx 11500$ main-sequence
stars with companions beyond $5000\au$ ($N_{\rm wb,0} = 287.5$)~\cite{Shariat:2025dxs},
distributed as $dN/ds \propto s^{-1.6}$
\cite{Yoo:2003fr,10.1093/mnrasl/sly206}, with a typical binary mass
$m_b = 0.6\Msun$. Normalizing over $[5000\au, \infty)$,  the initial number expected in the
counting window is
\begin{equation}
N_0(9\text{--}16\,{\rm kau})
 = N_{\rm wb,0}\,
   \frac{s_1^{-0.6} - s_2^{-0.6}}{(5000\au)^{-0.6}}
 = 59.0,
\label{eq:n0}
\end{equation}
with $s_1 = 9000\au$ and $s_2 = 16000\au$.
We adopt the same environmental inputs as Shariat et al.~\cite{Shariat:2025dxs}:
the Bo\"otes~I dark-matter density
$\rho_{\rm DM}=0.158\Msun\,{\rm pc}^{-3}$~\cite{Hayashi:2022wnw},
the stellar velocity dispersion
$\sigma_\star=4.6\kms$~\cite{Jenkins:2020blc}, and, following
Ref.~\cite{Graham:2024hah}, a perturber velocity dispersion
$\sigma_{\rm p}=8.50\kms$ corresponding to a relative encounter speed
$\sigma_{\rm rel}=9.66\kms$. We take the system age to be
$\tage=13$~Gyr.
This assumes that the binaries have been evolving for the full age of the old stellar population in Bo\"otes~I. Their formation history is not fully understood: they may have formed at later times, and some systems may have been formed or dynamically modified through processes such as exchanges involving a third, possibly dark, body~\cite{Penarrubia:2016ltr}. We therefore adopt the stellar age as a physically motivated benchmark for this old system, while noting that the inferred constraints depend on the effective time over which dynamical perturbations have acted.

The ratio $\Nobs/N_0 \simeq 0.36$ is consistent with the finding that
stellar flybys alone remove $30$--$70\%$ of binaries in this separation
range \cite{Shariat:2025dxs}. Dark perturbers are then excluded when their
additional disruption would reduce the surviving population below the
observed number of candidates.

\subsection{Disruption channels for point-mass perturbers}
\label{sec:channels}

Repeated weak flybys random-walk the binary's internal energy. The
diffusive disruption time can be expressed as~\cite{Shariat:2025dxs, 2008gady.book.....B}
\begin{equation}
\tdis \;=\; \frac{m_b\, \sigma}{16 \pi\, G\, n\, m_p^2\, a\, \lnL}\;.
\label{eq:B1closed}
\end{equation}
The Coulomb logarithm is
$\Lambda = a \sigma^2 / [G(m_b/2 + m_p)]$~\cite{HamiltonShaunak24}, i.e.\ the ratio of the maximal
tidal impact parameter $b_{\max} = a$ to the strong-deflection radius
$b_{90} = G(m_b/2 + m_p)/\sigma^2$~\cite{2008gady.book.....B}. For compact perturbers of mass $m_p=M$ at
fraction $\fdm$ of the DM, we have 
$n = \fdm \rho_{\rm DM}/M$.

A second, rarer channel comes from individual close encounters. A single
flyby passing sufficiently near the binary can inject enough energy to
unbind it in one event, rather than through the cumulative effect of many
weak encounters. Following
\cite{Shariat:2025dxs}, a
catastrophic timescale $\tcat = k_{\rm cat} (G\rho_{\rm DM})^{-1}
(G m_b/a^3)^{1/2}$ with $k_{\rm cat} = 0.07$~\cite{2008gady.book.....B} would have to be added for binaries wider
than the fringe scale $a_{\rm fringe} \propto \sigma m_b /(n m_p^2 t_{\rm age})$~\cite{HamiltonShaunak24, Penarrubia2019art}.
The total survival probability is
$\Psurv = \exp(-\tage/t_{\rm dis,total})$ with
$t_{\rm dis,total}^{-1} = \tdis^{-1} +
\Theta(a - a_{\rm fringe})\, \tcat^{-1}$.
For the Bo\"otes~I separations and dark-matter fractions relevant to our
limits, the catastrophic term does not turn on since
$a_{\rm fringe}\gg 16000\au$. The survival probability is therefore
controlled by cumulative weak encounters. This is the channel we
generalize to extended perturbers in Sec.~\ref{sec:constraints}.

The predicted number of surviving binaries in the counting window is
\begin{align}
\Npred(M, \fdm) &=  \nonumber \\
N_{\rm wb,0}
& \int_{9000\au}^{16{,}000\au} p(s)\,
\Psurv\big(a = s;\, M, \fdm\big)\, ds ,
\label{eq:npred}
\end{align}
with $p(s) \propto s^{-1.6}$ normalized over $[5000\au, \infty)$, the
survival evaluated at $a = s$, and stellar flybys deliberately excluded
from $\Psurv$. Both choices are conservative:
neglecting stellar flybys increases the predicted number of survivors,
while the $a\simeq s$ approximation affects the result only at the
$\lesssim 10\%$ level~\cite{Longhitano:2010zw}.
Under Poisson statistics, the $95\%$ exclusion is the locus
$\Npred(M, \fdm) = \mu_{\rm crit}$ with
$\mu_{\rm crit}(\Nobs{=}21) = 14.07$.

\section{Extension to finite-size perturbers}
\label{sec:formalism}

We work in the impulse approximation --- straight-line encounters at a
single effective speed $V = \sigma_{\rm rel}$ --- which is justified in
Bo\"otes~I because $\sigma_{\rm rel} = 9.66\kms$ far exceeds the binary
orbital speed ($v_{\rm orb} \lesssim 0.3\kms$ at $a = 10^4\au$), the same
argument used in the point-mass analysis
\cite{Shariat:2025dxs,HamiltonShaunak24}.
Ref.~\cite{Caputo:2026fuj}
provides a more general Fokker--Planck treatment of binary evolution under
stochastic perturbations. Here we consider the mean energy change induced
by individual encounters at fixed orbital parameters. For a spherical perturber
of arbitrary density profile, the transverse velocity kick on a star at
perpendicular distance $b$ from the trajectory is then~\cite{Pare:2026rtm}:
\begin{equation}
\Delta v_\perp(b) \;=\; \frac{2\, G\, \Mtd(b)}{b\, V}\,,
\label{eq:kick}
\end{equation}
where $\Mtd(b)$ is the perturber mass enclosed in the infinite cylinder
of radius $b$ around the trajectory. $\Mtd(R)$ can be computed from a 3D truncated profile $\rho(r)$ at $r_t$~\cite{WrightBrainerd2000, Pare:2026rtm}:
\begin{equation}
\Mtd(R) = \int_0^{r_t} 4\pi r^2 \rho(r)\; w(r, R)\, dr\,,
\label{eq:overlap}
\end{equation}
\begin{equation}
    w(r,R) = \begin{cases} 1\,, & r \leq R\,,\\[2pt]
1 - \sqrt{1 - R^2/r^2}\,, & r > R\,,\end{cases}
\end{equation}
where $w$ is the fraction of a thin spherical shell of radius $r$ that
lies inside the cylinder of radius $R$.

 Each star thus receives the transverse kick of Eq.~\eqref{eq:kick}
evaluated at its own perpendicular distance,
$|\Delta\mathbf v_i| = \Delta v_\perp(b_i)$, directed towards the
perturber trajectory; the internal orbit responds to
$\Delta\mathbf v_{\rm rel} = \Delta\mathbf v_1 - \Delta\mathbf v_2$,
which vanishes when both stars are kicked equally  and reduces to a single-star kick when one star
passes much closer to the trajectory than the other.
For an extended perturber we define
the reduced kick
\begin{equation}
\hat k(b)\equiv \frac{2G M_{\rm 2D}(b)}{b}
              = V\,\Delta v_\perp(b),
\label{eq:khat}
\end{equation}
so that the vector impulse on star \(i\) is
\begin{equation}
V\Delta \mathbf v_i
=
-\hat k(b_i)\,\hat{\mathbf b}_i ,
\end{equation}
where \(b_i\) is the perpendicular distance from star \(i\) to the
perturber trajectory and \(\hat{\mathbf b}_i\) points from the trajectory
towards the star. Hence
\begin{equation}
V\Delta \mathbf v_{\rm rel}
=
-\hat k(b_1)\hat{\mathbf b}_1
+
\hat k(b_2)\hat{\mathbf b}_2 .
\end{equation}
Squaring gives
\begin{equation}
|V\Delta \mathbf v_{\rm rel}|^2
=
\hat k(b_1)^2+\hat k(b_2)^2
-2\hat k(b_1)\hat k(b_2)
(\hat{\mathbf b}_1\cdot\hat{\mathbf b}_2).
\label{eq:dvrel_square}
\end{equation}
The first two terms are the two identical ``self'' contributions, while the
last term is the ``cross'' contribution that accounts for the cancellation
when the two stars receive similar kicks.
Integrating over the impact
parameter plane as in classic treatments of impulsive disruption by
point-mass perturbers
we get
\begin{equation}
\int d^2b\, |V\Delta \mathbf v_{\rm rel}|^2
=
2\hat S-2\hat C(d),
\label{eq:selfcross}
\end{equation}
with
\begin{equation}
\hat S
=
2\pi\int \hat k(b)^2\,b\,db,
\end{equation}
and
\begin{equation}
\hat C(d)
=
\int d^2b\,
\hat k(b_1)\hat k(b_2)
(\hat{\mathbf b}_1\cdot\hat{\mathbf b}_2).
\end{equation}
Here \(d\) is the projected separation of the two stars in the plane
perpendicular to the perturber trajectory. If the true binary separation is
\(a\) and the binary makes an angle \(\theta\) with the trajectory, then
\(d=a\sin\theta\). Only this projected separation enters the kick integral,
because the impulse depends on perpendicular distances to the trajectory.
Encounters with impact parameters in $d^2b$ occur at a rate $nV\,d^2b$,
and each changes the internal energy of the binary by
$\Delta E = \mu\,\mathbf v\cdot\Delta\mathbf v_{\rm rel}
+ \tfrac{1}{2}\mu\,|\Delta\mathbf v_{\rm rel}|^2$, with
$\mu = m_b/4$ the reduced mass; the term linear in the kick averages
to zero over isotropic orientations and orbital phases, so the mean
heating rate is
$\dot E = nV\!\int\! d^2b\;\tfrac{1}{2}\mu\,
\langle|\Delta\mathbf v_{\rm rel}|^2\rangle$.
We average over isotropic orientations of the binaries, and define the orientation-averaged heating integral
\[
J(a) \equiv \frac{1}{8}
\left\langle 2\hat S-2\hat C(a\sin\theta)\right\rangle_\theta ,
\]
so that, using Eq.~\eqref{eq:selfcross},
\[
\dot E = \frac{4 n \mu J(a)}{V}.
\]
The prefactor $1/8$ is fixed so that in the
point-mass limit the disruption time reproduces
Eq.~\eqref{eq:B1closed}, with the Coulomb logarithm replaced
by its orientation average,
$\lnL \to \langle \ln(a\sin\theta/b_{90})\rangle_\theta
= \lnL + \ln 2 - 1$.
Using \(|E|=Gm_b\mu/(2a)\), the diffusive disruption time becomes
\begin{equation}
t_{\rm dis}^{\rm ext}=\frac{|E|}{\dot E}
=\frac{Gm_b V}{8 n a J(a)} .
\label{eq:tdis}
\end{equation}
The point-mass heating integral is logarithmically divergent at small
impact parameters. To exclude the strong-scattering regime in a way
that also applies to extended perturbers, we adopt the following
 estimate
\begin{equation}
b_{\rm cut} = \frac{G\left[m_b/2 + \Mtd(b_{\rm cut})\right]}{\sigma^2}\,,
\label{eq:bcut}
\end{equation}
which reduces exactly to $b_{90}$ for a point mass, while for extended objects
with characteristic radius $\gg b_{90}$ the enclosed mass at
small $b$ is negligible, $b_{\rm cut}$ collapses to the quantity
$G(m_b/2)/\sigma^2 \approx 3 \, \rm{au}$ for our benchmark values, and the profile itself regularizes the integral.

For perturbers of
mass $M$ constituting a fraction $\fdm$ of the local dark-matter
density, the number density is $n = \fdm\,\rho_{\rm DM}/M$, the survival
probability of a binary of separation $a$ over the age of the system is
$\Psurv = \exp(-\tage/\tdis^{\rm ext})$ with $\tdis^{\rm ext}$ from
Eq.~\eqref{eq:tdis}, and the predicted number of surviving binaries
follows from Eq.~\eqref{eq:npred} exactly as in the point-mass analysis
of Sec.~\ref{sec:pointmass}. We evaluate these expressions for perturbers in the range $0.03$--$3000\Msun$, above which the spatial redistribution of massive clumps within the UFD may become important, and these effects are not considered here \cite{Graham:2024hah}.

In the point-mass limit our impulse calculation reproduces the standard
diffusive disruption rate up to a small Coulomb-logarithm convention
offset associated with the orientation average of the projected binary
separation.
We use the compact limit of this
calculation as the point-mass reference for finite-size effects. The imprint of the perturber's internal structure is expressed by normalizing its heating integral to that of a point mass
of equal total mass,
\begin{equation}
R(a) \;\equiv\; \frac{J_{\rm ext}(a)}{J_{\rm point}(a)}\,,
\label{eq:Rdef}
\end{equation}
which measures the suppression of the diffusive heating due to finite
size. A single sufficiently close flyby can also unbind a binary outright, as described in
Sec.~\ref{sec:channels}. We omit the catastrophic channel from our constraints, which can only underestimate the total
disruption rate, making our limits conservative.

\begin{figure*}[t]
\centering
\includegraphics[width=1.0\textwidth]{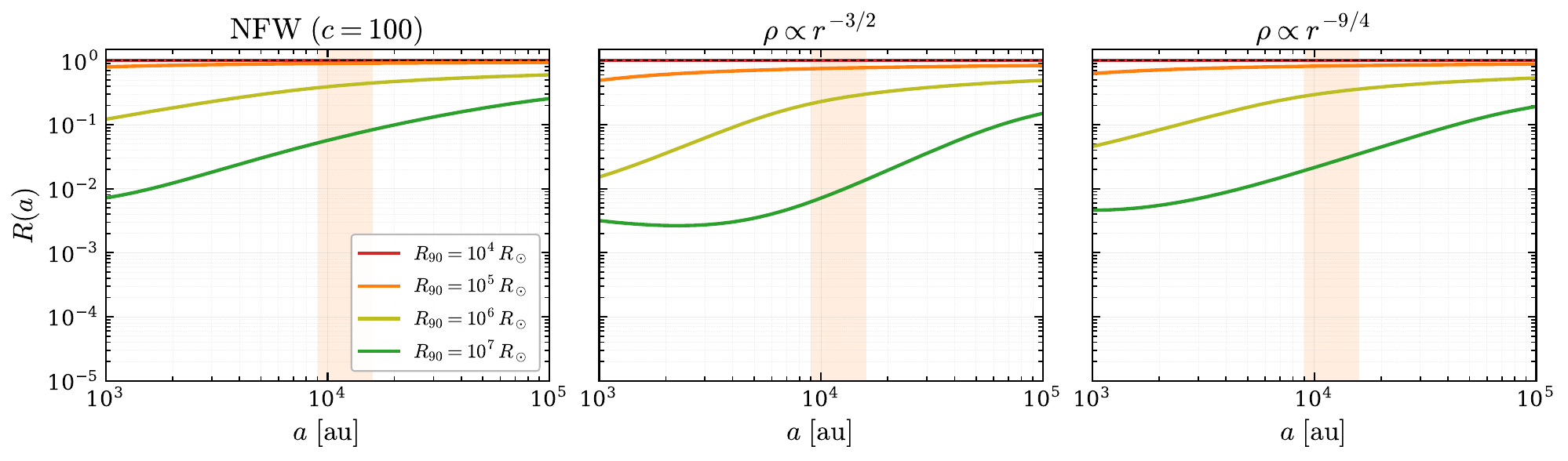}
\caption{Suppression of the diffusive wide-binary heating rate,
$R(a) = J_{\rm ext}/J_{\rm point}$, for extended perturbers of
$M = 10\Msun$ with fixed $\Rninety = 10^4$--$10^7\Rsun$ (one panel per
profile). The shaded band marks the Bo\"otes~I counting window
($9$--$16$~kau). Objects smaller than the strong-deflection radius
$b_{90} \simeq 2\times 10^4 \Rsun$ are exactly point-like ($R = 1$);
suppression sets in when $\Rninety$ exceeds $b_{90}$, deepens
monotonically with size, and is relieved with increasing $a$ as the
binary samples the perturber out to larger impact parameters.}
\label{fig:supp}
\end{figure*}

\section{Extended dark matter object constraints}
\label{sec:constraints}

To constrain objects of unknown internal structure without committing to
a formation scenario, we parameterize EDOs by two numbers: the total
mass $M$ and the radius $\Rninety$ that encloses $90\%$ of it,
$M({<}\Rninety) = 0.9\, M$, as it was done in
Ref.~\cite{Croon:2025yfj, Bringmann:2025cht, Croon:2024jhd}.
At fixed $(M, \Rninety)$ the remaining freedom is the shape of the
density profile, for which we adopt three spherical benchmarks that
bracket the concentrations expected for dark-matter substructure:  the NFW profile,
$\rho \propto [(r/r_s)(1 + r/r_s)^2]^{-1}$; a prompt-cusp--like power
law, $\rho \propto r^{-3/2}$ \cite{DelosWhite2023}, which is also the
inner UCMH profile adopted in Sec.~\ref{sec:ucmh_PR}; and the power law, $\rho \propto r^{-9/4}$. Each profile
must be truncated at some outer radius $r_t$, since the enclosed mass would grow without bound otherwise. For a power law
$\rho \propto r^{-\gamma}$ the enclosed mass scales as
$M({<}r) \propto r^{3-\gamma}$, so the requirement
$M({<}\Rninety) = 0.9\,M({<}r_t)$ fixes
\begin{equation}
r_t = 0.9^{-1/(3-\gamma)}\, \Rninety \,,
\end{equation}
giving $r_t = 1.073\,\Rninety$ for $\gamma = 3/2$ and
$r_t = 1.151\,\Rninety$ for $\gamma = 9/4$; the central normalization
then follows from $M({<}r_t) = M$.

The NFW profile carries two length scales, so one shape convention is
required: we fix the truncation concentration $c \equiv r_t/r_s = 100$, for which $\Rninety \approx 0.69\,r_t$ (as in Ref.~\cite{Croon:2024jhd}). The $\Rninety$ condition then fixes
$\Rninety/r_t = 0.690$, i.e.\ $r_s = 0.0145\,\Rninety$.
The benchmark radii $\Rninety = 10^4, 10^5, 10^6, 10^7\Rsun$
($= 46.5$, $465$, $4650$, $46{,}504\au$) bracket the Bo\"otes~I
counting window ($9$--$16$~kau) from well below to well above.

The profile influences the dynamics through the projected enclosed
mass $\Mtd(b)$ of Eq.~\eqref{eq:overlap}. By Eq.~\eqref{eq:kick}, a
star at perpendicular distance $b$ from the trajectory receives the
transverse impulse $\Delta v_\perp = 2 G \Mtd(b)/(b V)$ --- exactly the
kick of a point mass $\Mtd(b)$, because mass outside the impact
cylinder exerts no net transverse impulse. 

Its net effect is summarized by the suppression factor
$R(a)$ in Eq.~\eqref{eq:Rdef}: the fraction of the point-mass heating
rate that a binary of separation $a$ actually receives.

Fig.~\ref{fig:supp} shows $R(a)$ for $M = 10\Msun$ perturbers of the
four benchmark radii and the three benchmark profiles. Two comparison
scales organize its behaviour. The first is the strong-deflection
radius $b_{90} = G(m_b/2 + M)/\sigma_{\rm rel}^2 \simeq 2.0\times
10^3\, [(m_b/2 + M)/\Msun]\,\Rsun$, i.e.\ $\simeq 2\times 10^4\Rsun$
($98\au$) at $M = 10\Msun$: encounters inside $b_{90}$ are excluded
from the diffusive integral for extended and point-like perturbers, so any structure on scales $\Rninety < b_{90}$ is dynamically
invisible and $R = 1$ exactly. This is why the $\Rninety = 10^4\Rsun$
curves are flat at unity --- objects that compact are practically indistinguishable
from black holes of the same mass. The second scale is
the binary separation itself. Once $\Rninety > b_{90}$, encounters
with $b \lesssim \Rninety$ feel only the enclosed mass
$\Mtd(b) < M$, and the two stars receive increasingly similar kicks
that cancel in the difference; both effects suppress the heating, and
more strongly the larger $\Rninety$. The suppression is relieved as
$a$ grows: the heating integral is dominated by impact parameters up
to the binary separation, so wider binaries sample the perturber out
to larger $b$, enclosing more of its mass --- every curve therefore
rises with $a$ and tends to $R = 1$ when $a \gg \Rninety$. The shaded
band marks the Bo\"otes~I counting window.

The exclusion curves follow by the same statistical procedure as in
the point-mass case: for each profile and benchmark radius, the $95\%$
limit $\flim(M)$ solves $\Npred(M, \fdm) = \mu_{\rm crit}$
(Sec.~\ref{sec:channels}), with the survival probability set by the
absolute weak-encounter rate of Sec.~\ref{sec:formalism} alone (no
catastrophic contribution). By construction the
extended curves merge with the point-mass reference wherever
$b_{90}(M) \gtrsim \Rninety$, and they converge to the
point-mass limit for both profiles.

\begin{figure*}[t]
\centering
\includegraphics[width=0.99\textwidth]{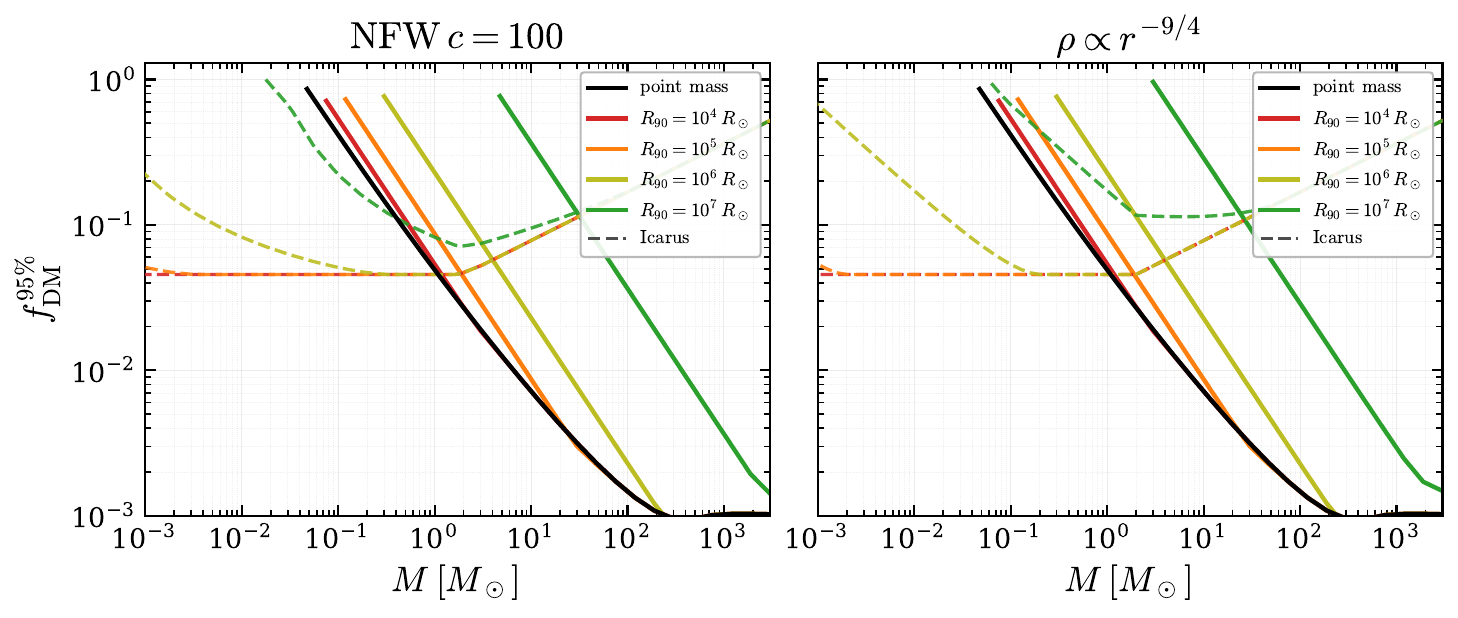}
\caption{$95\%$ exclusion on the dark-matter fraction $\fdm$ in
extended dark objects of mass $M$ and radius $\Rninety$ from the
survival of wide binaries in Bo\"otes~I (solid curves), for the two
benchmark profiles: NFW with $c = 100$ (left; the truncation
convention of Ref.~\cite{Croon:2024jhd}) and
$\rho \propto r^{-9/4}$ (right). Colors denote the benchmark radii
$\Rninety = 10^4$--$10^7\Rsun$; the black curve is the point-mass
limit of the same calculation. All wide-binary curves include the
diffusive channel only and are therefore conservative
(Sec.~\ref{sec:constraints}). Dashed curves show, with the same color
coding, the constraints from the Icarus caustic-crossing
event \cite{Croon:2025yfj}.}
\label{fig:flimgrid}
\end{figure*}

Fig.~\ref{fig:flimgrid} shows the resulting exclusion curves for the NFW and the $\propto r^{-9/4}$ profiles discussed above. Their
shape follows directly from the suppression physics: each curve
tracks the point-mass limit as long as $b_{90}(M) \gtrsim \Rninety$,
begins to deviate from it once the object is resolved, and is displaced
upward by an amount that grows with $\Rninety$. For
$\Rninety \leq 10^5\Rsun$ the limits are essentially point-like over
the whole mass range; at $10^6\Rsun$ they weaken by a factor of a few;
and at $10^7\Rsun$ --- objects larger than the binary separations ---
the exclusion onset is pushed up to a few solar masses and the limit
degrades by one to two orders of magnitude. At high mass all curves
deepen to $\fdm \sim 10^{-3}$.

The dashed curves show the corresponding constraints from the Icarus
caustic-crossing event \cite{Croon:2025yfj}, computed for the
same profiles, radii, and $\Rninety$ convention. The two probes are
 complementary in mass while overlapping in radius. Lensing
sensitivity is set by the effective Einstein radius, so Icarus
constrains objects far below a solar mass --- where wide binaries have
no sensitivity at all --- but its limits saturate at around
$\fdm \approx 5\times 10^{-2}$. The wide-binary limits switch on
near $\sim 0.1\Msun$ and strengthen monotonically with mass, crossing
below the Icarus bounds at $M \sim 1$--$2\Msun$ and reaching
$\fdm \sim 10^{-3}$, two orders of magnitude deeper, at
$M \gtrsim 10^2\Msun$. The combination of the two
probes therefore can exclude $\fdm = 1$ over many decades in
mass for objects as large as $\Rninety = 10^7\Rsun$, with wide binaries setting the constraint above a few solar masses.

\section{Ultracompact minihalos and limits on the primordial power
spectrum}
\label{sec:ucmh_PR}

The constraints of Sec.~\ref{sec:constraints} treat every point of the
$(M, \Rninety)$ plane as an independent hypothesis. A formation
scenario selects a one-dimensional track in this plane and thereby
inherits a definite abundance limit, which can be traced back to the
primordial conditions that produced the objects. In this section we
carry this out for UCMHs: dark-matter halos
formed from the collapse of large primordial overdensities that
entered the horizon deep in the radiation era. Following Refs.
\cite{Bringmann:2011ut, Bringmann:2025cht}, we consider UCMHs that have fully collapsed by
redshift $z_c = 1200$; such early-forming halos would develop the steep inner density profile $\rho \propto r^{-3/2}$ characteristic of
prompt cusps \cite{DelosWhite2023, Delos:2019mxl}, which we model by the same
construction as in Sec.~\ref{sec:constraints}: a truncated power law with
$M({<}r) \propto r^{3/2}$, outer radius
$r_t = 0.9^{-2/3}\,\Rninety = 1.073\,\Rninety$, and central
normalization fixed by $M({<}r_t) = M$.

We assume Gaussian primordial perturbations, a locally scale-invariant primordial power spectrum $\PR(k)$, and no post-formation mass growth, so that the present-day UCMH abundance is $\fdm=\beta$. At horizon crossing, the dark-matter density variance is then $\sigma^2_{\chi,H}\simeq0.91 \PR(k)$~\cite{Bringmann:2025cht}.
Finally, we assume that every UCMH that forms survives to
$z = 0$ and populates Bo\"otes~I at the mean dark-matter density.
Because late-time disruption would reduce the surviving minihalo population, neglecting it makes the resulting bound optimistic.

We note that alternative approaches to the power-spectrum recast are also possible, with Ref.\cite{Pare:2026rtm} inferring the full object population directly from the cosmological parameters, rather than reconstructing it from an abundance limit as it was done here following Refs.\cite{Bringmann:2011ut,Bringmann:2025cht}, and including an approximate treatment of tidal effects to estimate the present-day sizes and tidal radii of the objects.

The present approach nevertheless involves two main limitations. First, the abundance calculation assumes a one-to-one correspondence between each mode $k$ and a single halo mass, and does not remove overdense regions that lie inside larger collapsing regions, i.e.~it neglects the cloud-in-cloud problem~\cite{Bond:1990iw}. It is therefore best suited to power spectra with a single, narrow enhancement, which is also the regime in which UCMH-type treatments are usually applied. Second, the truncated $\rho\propto r^{-3/2}$ profile is fixed through the \((M,\Rninety)\) parametrisation of Sec.\ref{sec:constraints}, rather than derived from a formation model calibrated on simulations. For narrow/bumpy spectra, such a calibration is available: collapsed objects can be counted as density peaks according to their collapse time and assigned the $\rho\propto r^{-3/2}$ inner profile found in simulations \cite{Delos:2018ueo}. Implementing this mapping would provide a better-motivated normalization of the bound. The resulting constraints, however, depend only weakly on the assumed profile shape, as illustrated by the NFW, $\rho\propto r^{-3/2}$, and $\rho\propto r^{-9/4}$ results in Sec.\ref{sec:constraints}. The main uncertainty in the recast therefore enters through the mass–radius relation.

\begin{figure}[t]
\centering
\includegraphics[width=0.95\columnwidth]{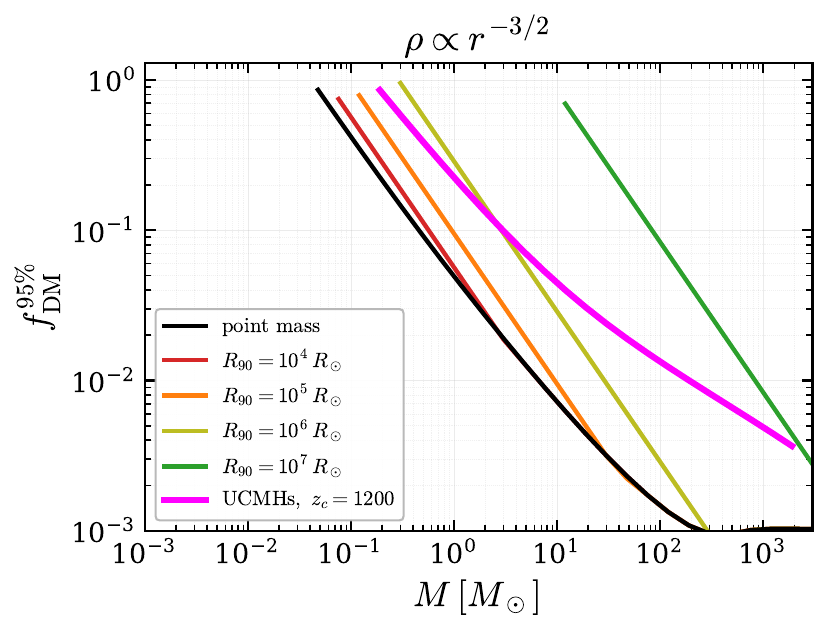}
\caption{$95\%$ exclusion on the dark-matter fraction in objects with
the prompt-cusp--like profile $\rho \propto r^{-3/2}$, for the four
benchmark radii $\Rninety = 10^4$--$10^7\Rsun$ (colored curves) and
the point-mass limit of the same calculation (black). The thick magenta curve shows the limit evaluated along the
physical UCMH track of Eq.~\eqref{eq:r90ucmh}, $\Rninety(M)$ for
$z_c = 1200$, which runs diagonally across the fixed-$\Rninety$ family,
from $\Rninety \simeq 4\times 10^5\Rsun$ at the low-mass end of the
excluded range to $\simeq 9\times 10^6\Rsun$ at the high-mass end.}
\label{fig:r32flim}
\end{figure}

We now apply the exclusion derived in Sec.\ref{sec:constraints} for the $\rho\propto r^{-3/2}$ profile (shown in Fig.\ref{fig:r32flim}) along the mass–radius relation predicted for the UCMH population. A UCMH of mass $M$ collapsing at $z_c$
has outer radius
$R(z_c) \simeq 21\,(1+z_c)^{-1} (M/\Msun)^{1/3}$~pc
\cite{RicottiOstrikerMack2008, Bringmann:2025cht}, and since
$M({<}r) \propto r^{3/2}$, the radius containing $90\%$ of its mass is
\begin{equation}
\Rninety = 0.9^{2/3}\, R(z_c)
= 7.2\times 10^{5}\, \left(\frac{M}{\Msun}\right)^{1/3} \Rsun
\quad (z_c = 1200)\,.
\label{eq:r90ucmh}
\end{equation}
Eq.~\eqref{eq:r90ucmh} places UCMHs with $M \approx 0.03$--$1900\Msun$ within the radius range shown in Fig.~\ref{fig:r32flim}. Across this mass range, the minihalos are extended enough for finite-size effects to matter, but still compact enough to be constrained.
Using this mass–radius relation, we compute the $95\%$ exclusion directly at the
physical radius (thick magenta curve in Fig.~\ref{fig:r32flim})
--- for each mass, the
survival probability is evaluated with the weak-encounter rate of
Sec.~\ref{sec:constraints} for the $\rho \propto r^{-3/2}$ profile of size
$\Rninety(M)$, and $\flim(M)$ solves
$\Npred(M, \fdm) = \mu_{\rm crit}$ as in Sec.~\ref{sec:channels}. The
resulting limit spans $M \approx 0.19$--$1893\Msun$, reaching
$\flim \approx 4.5\times 10^{-2}$ at $10\Msun$ and
$\approx 3.7\times 10^{-3}$ at the high-mass end of the computed
range. Identifying the abundance limit with the collapsed fraction gives
\begin{equation}
\beta_{\rm lim}(M) = \flim\big(M, \Rninety(M)\big)\,,
\label{eq:beta}
\end{equation}
We only translate this into a bound on the primordial spectrum when $\beta_{\rm lim}<0.5$, since larger values do not meaningfully constrain the Gaussian tail.

Each UCMH mass is identified with the dark-matter mass contained
in its comoving scale $k$~\cite{Bringmann:2025cht},
\begin{equation}
M(k) = \frac{4\pi}{3}\, \frac{\rho_{\chi,0}}{k^3}
= 1.39\times 10^{11} \left(\frac{k}{{\rm Mpc}^{-1}}\right)^{-3}
\Msun\,,
\label{eq:mofk}
\end{equation}
with $\Omega_\chi = 0.264$ and $h = 0.674$. 
For Gaussian perturbations, the fraction of horizon patches that collapse by $z_c$ is determined by the high-density tail of the density-contrast distribution~\cite{Bringmann:2011ut,Bringmann:2025cht},
\begin{equation}
\beta(k) = \frac{1}{2}\,
{\rm erfc}\!\left[\frac{\dmin(k)}{\sqrt{2}\,\sigma_{\chi,H}(k)}\right],
\end{equation}
where $\dmin(k)$ is the critical dark-matter density contrast at
horizon entry required for collapse by $z_c$ (the
$\delta_{\max} \sim \mathcal{O}(1)$ upper cutoff on collapsing
perturbations is exponentially subleading and neglected). Because
$\beta$ depends exponentially on $\dmin^2/\sigma^2_{\chi,H}$, even the
moderate abundance limits $\beta_{\rm lim} \sim 10^{-3}$--$10^{-1}$
obtained here translate into tight limits on the variance, and hence
on the spectrum. Since $\beta$ is monotonic in $\sigma_{\chi,H}$, the abundance limit
inverts directly: using $\sigma^2_{\chi,H} = 0.91\,\PR$
\cite{Bringmann:2025cht},
\begin{equation}
\PR^{\rm lim}(k) = \frac{\dmin(k)^2}
{2 \times 0.91 \times
\left[{\rm erfc}^{-1}\!\big(2\beta_{\rm lim}(k)\big)\right]^2}\,.
\label{eq:prlim}
\end{equation}
The critical contrast $\dmin(k)$ is the minimum density contrast at
horizon entry required for collapse by $z_c = 1200$: collapse occurs once the linearly grown
contrast reaches the standard threshold
$\delta_c \simeq 1.686$, so $\dmin(k)$ scales
inversely with the linear growth factor accumulated between horizon
entry and $z_c$. Modes with larger $k$ enter the horizon earlier and, while inside the horizon during
radiation domination, dark-matter perturbations grow logarithmically
with the scale factor~\cite{}; the growth after
equality is the same for all $k$. The threshold therefore falls
logarithmically with $k$,
\begin{equation}
\dmin(k) = \dmin(k_0)
\left[1 + \frac{\ln(k/k_0)}{c_\delta}\right]^{-1}\,,
\end{equation}
where we take $k_0 = 3\,{\rm Mpc}^{-1}$ with
$\dmin(k_0) = 1.03\times 10^{-2}$ from
Ref.~\cite{Bringmann:2025cht}, and $c_\delta = 2.911$ is fixed by
their second quoted value, $\dmin = 2.25\times 10^{-3}$ at
$k = 10^5\,{\rm Mpc}^{-1}$. Since $\PR^{\rm lim} \propto \dmin^2$,
the residual freedom in this one-parameter form affects the
normalization of the bound at the $\sim 10\%$ level.

\begin{figure*}[t]
\centering
\includegraphics[width=1.5\columnwidth]{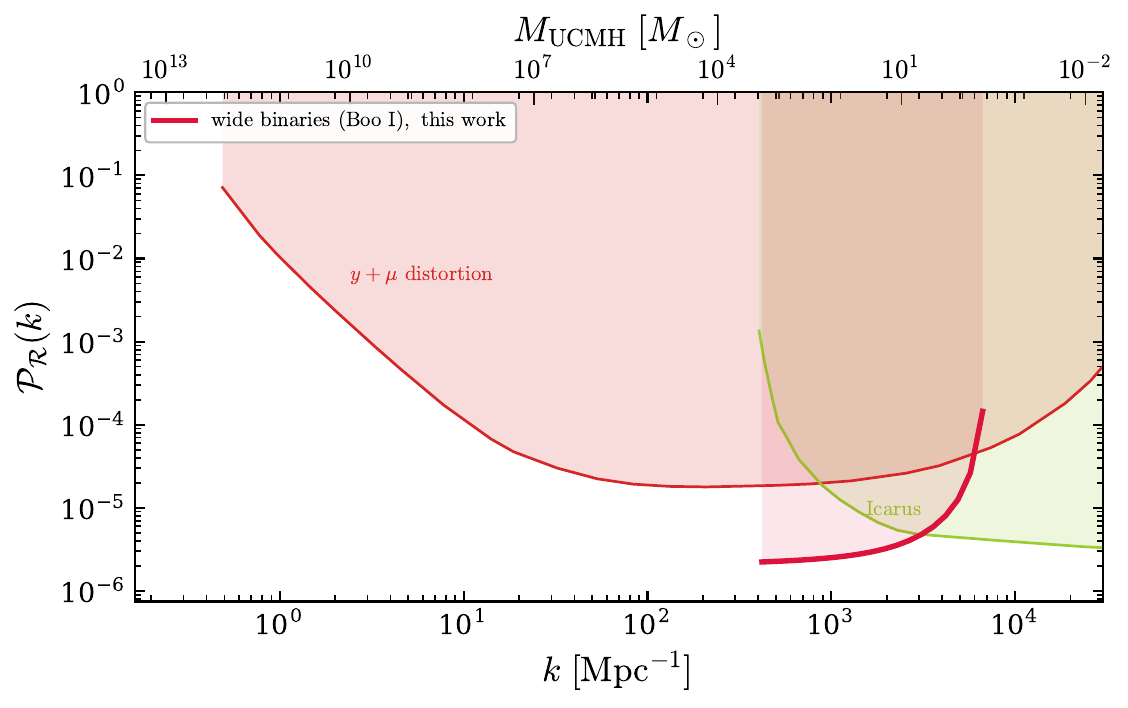}
\caption{Illustrative $95\%$ upper limit on the primordial curvature
power spectrum from UCMH disruption of wide binaries in Bo\"otes~I
(red curve and shaded region), under the assumptions stated in the
text: Gaussian perturbations, locally scale-invariant $\PR$, collapse
by $z_c = 1200$, a truncated $\rho \propto r^{-3/2}$ inner profile, no
post-formation mass growth, no survival correction, and the
monochromatic mass--scale mapping of Eq.~\eqref{eq:mofk}. The low-$k$ termination reflects the largest
radius computed.}
\label{fig:prcomp}
\end{figure*}

The resulting limit is shown in Fig.~\ref{fig:prcomp}, overlaid on the
existing small-scale bounds shown in \cite{Bringmann:2025cht}. It reaches
$\PR^{\rm lim} \approx 2.2\times 10^{-6}$ at the low-$k$ end of the band
($k \approx 4\times 10^{2}\,{\rm Mpc}^{-1}$, corresponding to the highest
UCMH masses $M \sim 10^{2}$--$10^{3}\Msun$) and rises steeply toward the
high-$k$ exclusion onset at $k \approx 6.6\times 10^{3}\,{\rm Mpc}^{-1}$.  These scales lie far beyond those directly probed by the CMB, which constrains \(\PR(k)\) over much larger cosmological scales, where
$\PR \approx 2\times 10^{-9}$.

Over the range $k \simeq 4\times 10^{2}$--$3\times10^{3}\,{\rm Mpc}^{-1}$,
roughly a decade, the bound is competitive with the strongest constraints
available at these scales: it lies below the CMB $y$/$\mu$-distortion
limit~\cite{Chluba:2012we} and the low-$k$ portion of the Icarus lensing
constraint. Toward the high-$k$ end of the range, however, the constraint
weakens: as the UCMH mass falls below $\sim 1\Msun$, the abundance limit
$\flim$ becomes less restrictive and $\PR^{\rm lim}$ increases, while the
Icarus caustic-crossing constraint continues to improve, reaching a few
$\times 10^{-6}$ near $k\sim10^{4}$--$10^{5}\,{\rm Mpc}^{-1}$.

This result should be interpreted as an estimate of the potential reach
of ultra-faint-dwarf wide binaries as a probe of the primordial spectrum.
The recast assumes Gaussian perturbations, a locally scale-invariant
$\PR$, collapse by $z_c=1200$, a truncated $\rho\propto r^{-3/2}$ profile,
and no mass growth after formation. It also neglects late-time disruption
of the minihalos, which would reduce the surviving perturber population,
and inherits the uncertainties associated with the dynamical modelling in
Sec.~\ref{sec:constraints} as well as with the mapping between the
primordial spectrum and the UCMH population.

\section{Conclusions}
\label{sec:conclusions}

We have generalized the constraints derived from the survival of wide binaries in Bo\"otes~I from point-like perturbers to extended dark objects with arbitrary spherical density profiles. The key ingredient is the projected-mass impulse, in which
the point-mass kick $2Gm_p/(bV)$ is replaced by $2G\Mtd(b)/(bV)$, with
$\Mtd(b)$ the mass enclosed in the impact cylinder; in the compact-object limit, the calculation converges to the Bo\"otes~I point-mass constraint of Ref.~\cite{Shariat:2025dxs}.

Finite size enters through two characteristic scales. The first one is the strong-deflection
radius $b_{90}$: objects with $\Rninety \lesssim b_{90}$
disrupt binaries exactly like black holes of the same mass. Once
$\Rninety$ exceeds $b_{90}$, the degree of suppression is governed by
$\Rninety$ relative to the binary separation itself --- penetrating
encounters deliver reduced, partially cancelling kicks until
$a \gg \Rninety$. The resulting $95\%$ limits $\flim(M,\Rninety)$ stay
within a factor of a few of the point-mass bound for
$\Rninety \lesssim 10^{5}\Rsun$, deepen to $\fdm \sim 10^{-3}$ at high
mass, and weaken by one to two orders of magnitude only for the largest
objects considered ($\Rninety = 10^{7}\Rsun$).

Mapping the $\rho \propto r^{-3/2}$ results onto ultracompact minihalos
collapsing at $z_c = 1200$ converts the dynamical bound into an
illustrative, purely gravitational constraint on the primordial curvature
power spectrum, $\PR \lesssim 5\times 10^{-6}$ over $k \approx 4\times 10^{2}$--$3\times 10^{3}\,{\rm Mpc}^{-1}$, competitive
with the strongest existing small-scale bounds where they overlap.

The more complete Fokker--Planck treatment of binary evolution developed in
Ref.~\cite{Caputo:2026fuj} provides a natural framework for future
refinements, including the Maxwellian velocity distribution and the full
evolution of the orbital parameters. Further extensions could incorporate
physical mass--size relations, extended mass functions, and a
simulation-calibrated mapping from the primordial power spectrum to the
halo population and density profiles \cite{Delos:2018ueo}. Applying these
methods to additional ultra-faint dwarfs and larger future catalogues from
Rubin/LSST and Roman will provide new opportunities to test massive dark
perturbers, although the ultimate sensitivity will depend on controlling
the astrophysical uncertainties associated with the binary population.

\section{Acknowledgements}
I am grateful to Julien Lavalle and M. Sten Delos for valuable comments on the draft.
I am supported by the STFC under grant ST/X000753/1. I also want to acknowledge support of the grant PID2024-161668NB-I00.

\bibliography{references}
\bibliographystyle{apsrev4-1}

\end{document}